\documentstyle[aps,preprint,array,12pt]{revtex}

\begin{document}
\begin{sl}
\title{NONLOCAL EFFECTS IN QUANTUM GRAVITY}
\author{ALI SHOJAI\footnote{Email: SHOJAI@NETWARE2.IPM.AC.IR}$^{,1,3}$\\and
\\FATIMAH SHOJAI\footnote{Email: FATIMAH@NETWARE2.IPM.AC.IR}$^{,3}$\\and
\\MEHDI GOLSHANI$^{2,3}$}
\address{$^1$Department of Physics, Tarbiat Moddarres University,\\P.O.Box 14155-4838, Tehran, IRAN.}
\address{$^2$Department of Physics, Sharif University of Technology,\\P.O.Box 9161, Tehran 11365, IRAN.}
\address{$^3$Institute for Studies in Theoretical Physics and Mathematics,
\\P.O.Box 19395-5531, Tehran, IRAN.}
\date{\today}
\maketitle
\begin{abstract}
{\it Recently\cite{BQG}, a new quantum gravity theory was presented in which
the quantum effects were represented by the conformal degree of freedom of 
the space--time metric. In this work we show that in the framework of this 
theory quantum gravity is nonlocal.}
\end{abstract}
\section{INTRODUCTION}
\par
In a recent work\cite{BQG}, it was shown that the quantal behaviour of matter
can be understood as a purely geomertical effect. In fact the conformal degree
of freedom of the space--time metric would be determined by quantal effects. 
In this view, the geometry has two physical significances. First, its 
conformal degree of freedom represents what is usually called {\it the 
quantal effects}. Secondly, its other degrees of freedom determine the causal 
structure of the space--time. This second part, in the absence of quantal 
effects (where the conformal factor is a constant), is called 
{\it classical gravity}. These two parts are highly coupled so that the 
theory is expected to be a{\it quantum gravity\/} theory. 

The theory, 
primarily rests on the de-Broglie--Bohm quantum theory\cite{Bohm}, which 
is the causal counterpart of quantum mechanics. In Bohmian mechanics, 
any particle is always accompanied by an objectively real field exerting 
some force on the particle. This is called the {\it quantum force}. 
In the case of 
relativistic particles, the quantum potential is nothing but the mass of 
the particle. So the equation of motion of a relativistic particle 
is\cite{BQG}:
\begin{equation}
\label{QNSL}
\frac{d({\cal M}u_{\mu})}{d\tau}=c^2\nabla_{\mu}{\cal M}
\end{equation}
where
\begin{equation}
\label{QMF}
{\cal M}^2=m^2+\frac{\hbar^2}{c^2}\frac{\Box |\Psi|}{|\Psi|}
\end{equation}
and
\begin{equation}
\Box \Psi+\frac{m^2c^2}{\hbar^2}\Psi=0
\end{equation}
The theory rests also on the de-Broglie {\it ansatz} that the presence of 
quantum force is identical to having a curved space--time. This fact can 
be seen simply by writing (\ref{QNSL}) in the Hamilton--Jacobi form:
\begin{equation}
\label{QHJE}
g^{\mu\nu}\nabla_{\mu}S\nabla_{\nu}S={\cal M}^2c^2;
\ \ \ \ \ \ \ \ \nabla_{\mu}S={\cal M}u_{\mu}
\end{equation}
Equation (\ref{QHJE}) can be rewritten as:
\begin{equation}
\tilde{g}^{\mu\nu}\tilde{\nabla}_{\mu}S\tilde{\nabla}_{\nu}S=m^2c^2;
\ \ \ \ \ \ \ \ \tilde{g}_{\mu\nu}=\frac{{\cal M}^2}{m^2}g_{\mu\nu}
\end{equation}
Accordingly, an appropriate action for quantum gravity is written in 
\cite{BQG}. The corresponding equations of motion are:
\begin{equation}
{\cal R}\Omega+6\Box\Omega+\frac{2\kappa}{m}\rho \Omega 
\left ( \nabla_{\mu}S\nabla^{\mu}S-2m^2\Omega^2\right )
+2\kappa \lambda\Omega=0
\end{equation}
\begin{equation}
\nabla_{\mu}\left (\rho \Omega^2 \nabla^{\mu}S \right )=0
\end{equation}
\begin{equation}
\left ( \nabla_{\mu}S \nabla^{\mu}S -m^2\Omega^2\right )
\Omega^2\sqrt{\rho}+\frac{\hbar^2}{2m}
\left [ \Box\left (\frac{\lambda}{\sqrt{\rho}}\right ) 
-\lambda\frac{\Box\sqrt{\rho}}{\rho}\right ]=0
\end{equation}
\[ {\cal G}_{\mu \nu}
-\frac{\left [ g_{\mu \nu}\Box -\nabla_{\mu}\nabla_{\nu}\right ]
\Omega^2}{\Omega^2} -6 \frac{\nabla_{\mu}\Omega \nabla_{\nu}\Omega}{\omega^2}
+3g_{\mu \nu}\frac{\nabla_{\alpha}\Omega \nabla^{\alpha}\Omega}{\Omega^2}
+\frac{2\kappa}{m}\rho \nabla_{\mu}S \nabla_{\nu}S-\frac{\kappa}{m}\rho 
g_{\mu \nu} \nabla_{\alpha}S \nabla^{\alpha}S \]
\begin{equation}
+\kappa m \rho \Omega^2 g_{\mu \nu}
+\frac{\kappa\hbar^2}{m^2}\left [ \nabla_{\mu}\sqrt{\rho}\nabla_{\nu}
\left ( \frac{\lambda}{\sqrt{\rho}}\right )
+\nabla_{\nu}\sqrt{\rho}\nabla_{\mu}\left ( \frac{\lambda}{\sqrt{\rho}}\right )\right]
-\frac{\kappa\hbar^2}{m^2}g_{\mu \nu}\nabla_{\alpha}\left (\lambda 
\frac{\nabla^{\alpha}\sqrt{\rho}}{\sqrt{\rho}}\right )=0
\end{equation}
\begin{equation}
\Omega^2=1+\frac{\hbar^2}{m^2}\frac{\Box\sqrt{\rho}}{\sqrt{\rho}}
\end{equation}
where $\Omega$ is the conformal degree of freedom of the metric, $\lambda$
is a lagrange multiplier and $\rho=\Psi^*\Psi$ is the matter density.
A special case is when $\lambda$ can be expanded in powers of 
$\alpha=\hbar^2/m^2$. Then, it can be simply shown \cite{BQG} that in this
case $\lambda=0$, and the equations of motion are:
\begin{equation}
\label{CCC}
\nabla_{\mu}\left (\rho \Omega^2 \nabla^{\mu}S \right )=0
\end{equation}
\begin{equation}
\label{AAA}
\nabla_{\mu}S \nabla^{\mu}S =m^2\Omega^2
\end{equation}
\begin{equation}
\label{DDD}
{\cal G}_{\mu \nu}=-\kappa {\cal T}^{(m)}_{\mu\nu}-\kappa{\cal T}^{(\Omega)}_{\mu\nu}
\end{equation}
\begin{equation}
\label{EEE}
{\cal T}^{(m)}_{\mu\nu}=\frac{\rho}{m}\nabla_\mu S\nabla_\nu S
\end{equation}
\begin{equation}
\label{FFF}
\kappa{\cal T}^{(\Omega)}_{\mu\nu}=\frac{\left [ g_{\mu \nu}\Box 
-\nabla_{\mu}\nabla_{\nu}\right ]
\Omega^2}{\Omega^2} +6 \frac{\nabla_{\mu}\Omega \nabla_{\nu}\Omega}{\omega^2}
-3g_{\mu \nu}\frac{\nabla_{\alpha}\Omega \nabla^{\alpha}\Omega}{\Omega^2}
\end{equation}
\begin{equation}
\label{BBB}
\Omega^2=1+\alpha\frac{\Box\sqrt{\rho}}{\sqrt{\rho}}
\end{equation}
As one can see, there are two contributions to the background metric
($g_{\mu\nu}$). First, we have ${\cal T}^{(m)}_{\mu \nu}$ which 
represents the 
gravitational effects of matter. Second, there is 
${\cal T}^{(\Omega)}_{\mu \nu}$ which is a result of the quantal effects
of matter. Since in the evaluation of the ${\cal T}^{(\Omega)}_{\mu\nu}$
the background metric is used, the gravitational and quantal contributions 
to the background metric are so highly coupled that no one without the 
other has any physical significance.
In this way the theory is a {\it quantum 
gravity\/} theory. 

It must be pointed out here that, since the conformal 
factor is meaningless as $\rho\rightarrow 0$, the geometry looses 
its meaning at this limit. This is a desired property, because it is in 
accord with Mach's principle, which states that for an empty universe 
the space--time should be meaningless. 

A special aspect of the quantum force 
is that it is highly nonlocal. This property, 
which can be seen from the equation 
(\ref{QMF}), is an experimental matter of fact \cite{ASP}. Since the mass 
field given by (\ref{QMF}) represents the conformal degree of freedom of 
the physical metric, quantum gravity is expected to be highly nonlocal. 
In the next section, this is shown explicitley for a specific problem. 
\section{Illustration of nonlocal effects in quantum gravity}
\par
In order to illustrate how nonlocal effects can appear in quantum gravity 
through quantum potential, suppose that matter distribution is localized 
and has spherical symmetry. Then, one has:
\begin{equation}
\rho=\rho(t;r)
\end{equation}
\begin{equation}
\Omega=\Omega(t;r)
\end{equation}
Suppose, furthermore, that matter is at rest:
\begin{equation}
-\nabla_0S=E(t;r)\ \ as\ \ r\rightarrow \infty
\end{equation}
\begin{equation}
\nabla_iS=0;\ \ \ \ i=1,2,3
\end{equation}
One expects that at large $r$, where there is no matter, the background 
metric would be of the Schwartzschield form:
\begin{equation}
\label{GGG}
g_{\mu\nu}=\left ( \begin{array}{cccc}1-r_s/r&0&0&0\\0&-1/(1-r_s/r)&0&0
\\0&0&-r^2&0\\0&0&0&-r^2\sin^2\theta\\
\end{array} \right )\ \ \ as\ \ r\rightarrow \infty
\end{equation}
where $r_s$ is a constant (the Schwartzschield radius). The validity of 
this approaximation will be examined at the end. The equation of motion 
(\ref{AAA}) relates $E$ and $\Omega$:
\begin{equation}
E=\frac{m\Omega}{\sqrt{1-r_s/r}}
\end{equation}
In order to calculate the conformal factor $\Omega$, one needs the specific 
form of $\rho$. It must be a localized function at $r=0$. So we choose it as:
\begin{equation}
\rho(t;r)=A^2\exp[-2\beta(t)r^2]
\end{equation}
Using the relation (\ref{BBB}), the conformal factor can be simply calculated.
This leads to:
\[ \Omega^2=1+\alpha[\dot{\beta}^2r^4-\ddot{\beta}r^2+4\beta^2r] \]
from which we get:
\begin{equation}
\Omega^2\simeq \alpha\dot{\beta}^2r^4\ \ \ as\ \ r\rightarrow \infty
\end{equation}
Now it is a simple task to examine that the continuity equation (\ref{CCC}) 
is satisfied automatically as $r\rightarrow \infty$. This solution 
is an acceptable one, only if the generalized Einstein's equations (\ref{DDD}) 
are satisfied. This is so if ${\cal T}^{(\Omega)}_{\mu\nu}\rightarrow 0$ as
$r\rightarrow\infty$. It can be shown that in the limit $r\rightarrow\infty$
we have:
\begin{equation}
\frac{\Box\Omega^2}{\Omega^2}=2(\ddot{\beta}/\dot{\beta})^2
+2\dot{\ddot{\beta}}/\dot{\beta}-20/r
\end{equation}
\begin{equation}
\frac{\nabla_0\nabla_0\Omega^2}{\Omega^2}=(\ddot{\beta}/\dot{\beta})^2
+\dot{\ddot{\beta}}/\dot{\beta}
\end{equation}
\begin{equation}
\frac{\nabla_1\nabla_1\Omega^2}{\Omega^2}=12/r^2
\end{equation}
\begin{equation}
\frac{\nabla_1\nabla_0\Omega^2}{\Omega^2}=(8\ddot{\beta}/r\dot{\beta})
\end{equation}
\begin{equation}
\left ( \frac{\nabla_0\Omega}{\Omega}\right )^2
=(\ddot{\beta}/\dot{\beta})^2
\end{equation}
\begin{equation}
\frac{\nabla_1\Omega\nabla_0\Omega}{\Omega^2}=(2\ddot{\beta}/r\dot{\beta})
\end{equation}
So provided that higher time derivatives of the scale factor of matter 
density 
($\beta$) are small with respect to its first time derivative, 
that is:
\begin{equation}
\frac{\ddot{\beta}}{\dot{\beta}}\simeq0; 
\frac{\dot{\ddot{\beta}}}{\dot{\beta}}\simeq 0\ \ and\ so\ on
\end{equation}
one has:
\begin{equation}
\lim_{r\rightarrow\infty}{\cal T}^{(\Omega)\nu}_{\mu}=0
\end{equation}
Also we have from (\ref{EEE}):
\begin{equation}
\lim_{r\rightarrow\infty}{\cal T}^{(m)\nu}_{\mu}=0
\end{equation}
So at large distances $g_{\mu\nu}$ satisfies Einstein's equations in vaccum, 
${\cal G}_{\mu\nu}=0$. Therefore, the solution (\ref{GGG}) is acceptable. In this 
way we find a solution to the quantum gravity equations at large distances.

Consequently, if the time variation of $\beta$ is small, the physical metric
$\tilde{g}_{\mu\nu}=\Omega^2g_{\mu\nu}$ is given by:
\begin{equation}
\label{HHH}
\lim_{r\rightarrow\infty}\tilde{g}_{\mu\nu}=\alpha\dot{\beta}^2r^4
g^{(Shwarzschield)}_{\mu\nu}
\end{equation}
An important points must be noted here. As it was shown, a change in 
matter distribution (due to $\dot{\beta}$) instantaneosely alters the 
physical metric. This is because of the appearance of $\dot{\beta}(t)$ in 
equation(\ref{HHH}) and it comes from the quantum potential term. 

We conclude 
that the specific form of the quantum potential leads to the appearance of 
nonlocal effects in quantum gravity.

\end{sl}
\end{document}